\documentclass[%
 reprint,
 amsmath,amssymb,
 aps,
]{revtex4-2}

\usepackage{color}
\usepackage{upgreek}
\usepackage{graphicx}
\usepackage{dcolumn}
\usepackage{bm}

\newcommand{\muz}{\upmu_0}

\begin{document}


\title{Axisymmetric Plasma Equilibria with Toroidal and Poloidal Velocity Fields:\\
Tokamak Relevant Configurations }

\author{Giovanni Montani$^{1,2}$}
\author{Matteo Del Prete$^2$}%
 \email{matteo.delprete@uniroma1.it}
\affiliation{$^1$ ENEA, Fusion and Nuclear Safety Department, C.R. Frascati\\
Via E. Fermi 45, 00044 Frascati (Roma), Italy%
}%
\affiliation{$^2$ Physics Department, ``Sapienza'' University of Rome,\\
P.le Aldo Moro 5, 00185 Roma, Italy%
}%

\date{\today}

\begin{abstract}
We analyze an axisymmetric equilibrium of a plasma endowed with toroidal and poloidal velocity fields, with the aim to characterize the influence of the global motion on the morphology of the magnetic confinement. We construct our configuration assuming that the poloidal velocity field is aligned with the poloidal magnetic field lines and, furthermore, we require that the plasma mass density depend on the magnetic flux function (or equivalently, that the plasma fluid be incompressible). We then derive a sort of Grad--Shafranov equation for such an equilibrium and implement it to tokamak relevant situations, with particular reference to TCV--like profiles. The main result of the present study concerns the emergence, in configurations associated to a double--null profile, of a closed surface of null pressure encorporating the two X-points of the magnetic configuration. This scenario suggests the possible existence of a new regime of the plasma equilibrium, corresponding to an improved plasma confinement near the X-points and a consequent reduced power transfer to the tokamak divertor.
\end{abstract}

\maketitle


\section{Introduction}
The possibility to confine a plasma into a finite region of space necessarily requires the application of external magnetic fields able to produce the requested balance of the fundamental forces acting on the plasma itself \cite{landau}. However, the achievement of a steady equilibrium for a fusion relevant laboratory plasma is far from being available in experimental operations, both for the limited time duration of the discharge and for the unavoidable plasma instabilities \cite{wesson,biskamp}. Therefore, in the context of tokamak machines we can consider as equilibrium a magnetic configuration which lives for a much longer time than the typical instability inverse rates and, on the other hand, sufficiently short to prevent the observation of significant changes in the global distribution of the magnetic and internal energy of the plasma.

Historically (see the original studies in \cite{grad,shafranov}) the basic magnestostatic equation underlying the equilibria consists of the balance between the pressure gradient and the $\vec{j}\wedge\vec{B}$ force, without considering any contribution due to velocity fields, and to the mass density as a consequence.
This scenario was appropriate for the pioneering experiments of the plasma confinement and also for many later machines, but, since the early nineties it became clear \cite{hassam93} that tokamak plasmas could be endowed with both toroidal and poloidal rotation fields (known as \emph{spontaneous rotation}), even reaching values comparable to the sound speed \cite{rice07}.

The necessity to include velocity fields in the description of a tokamak equilibrium is increasingly relevant to achieve a satisfactory level of predictivity. In the next generation large sized tokamaks like ITER \cite{iter18}, JT-60SA \cite{giruzzi19}, DTT \cite{dtt19}, but also in present medium sized tokamak experiments, for instance MAST-U \cite{harrison19} and TCV \cite{coda17}, the request to include toroidal and poloidal velocity fields is a consequence of the spontaneous rotation phenomenon, cited above, but also due to the interaction of the plasma with additional heating systems, in particular when injection of hot neutral beams is utilized. Furthermore, the operation of the current machines in the H-mode regime seems to be favoured by transport barriers due to the radial electric field which lives in the equilibrium by virtue of the presence of poloidal velocity fields \cite{burrell97}.

Studies to include the velocity contributions in tokamak equilibria are present in literature, see for instance \cite{maschke,ogilvie,guazzotto05,greci}, and are mainly focused on the possibility to include toroidal motions of the plasma, by making use of the concepts of generalized pressure or Bernoulli functions.
For more recent systematic approaches see \cite{farengo,guazzotto21}; moreover, a specific semi-analytical technique to investigate, on one hand the influence on equilibria of the resistive diffusion and, on the other hand of the toroidal field velocity, has been considered in \cite{egse} and \cite{dpmarxiv}, respectively.
In the present letter, we construct an equilibrium configuration of a plasma endowed simultaneously with toroidal and poloidal velocities, arriving to a specific reformulation of the standard Grad--Shafranov equation.
We base our analysis on two main assumptions: i) the poloidal velocity field follows the corresponding poloidal magnetic lines; ii) the plasma mass density depends on the magnetic flux function, or equivalently, the fluid can be regarded as incompressible.

The implementation of the obtained Grad--Shafranov-like equation has the peculiarity to be nonlinear in the poloidal magnetic field component, because the magnetic pressure gradient remain, together with the thermodynamical pressuire gradient, into the force balance fixing the equilibrium. The implementation to the tokamak equilibrium configuration immediately outline the expected deviation of the magnetic surfaces form the isobar ones. As a consequence, three different regions of the plasma confinament can be identified: i) a region which is within the separatrix and the isobar of zero pressure (the most confined region); ii) the region which is out of the isobar of vanishing pressure, but which is still within the separatrix (a lower confinement condition); iii) the usual scrape--off layer region (there no real confinement exists).

However, in general, the separatrix and the zero pressure region intersect each other and therefore more hybrid situations can take place, like the possibility for the plasma costituents to cross the isobar of vanishing pressure along the magnetic surfaces and therefore, in principle, due to parallel transport. However, the most interesting result of our analysis emerges when we implement the present scenario to a TCV-like configuration in the presence of a magnetic profile endowed with two (up and down) X-points \cite{tcvlike}. We constructed an equilibrium for such a double null profile and observed the formation of a closed line of zero pressure which encircles the two X-points. The presence of these two plasma lobi suggests that, in the presence of velocity fields, the plasma around the null is better confined with respect to the absence of any motion. This has reasonable implications on the heat and particle transport form the X-point towards the divertor. By other words, the "private region'' (\emph{i.e.} the region enclosed by the divertor surface and the magnetic configuration outer legs) results divided into two parts: one inside a lobe of zero pressure and one at all equivalent to that region commonly observed in the scrape--off layer. The existence of these symmetric (up and down) minor lobi can have an important role in protecting the divertor from excessive power tranfers from the plasma and it should be further investigated under many points of view.

\section{Basic equations}\label{sec1}
We consider the equations describing a steady ideal MHD fluid, characterized by a mass density $\rho$, a pressure $p$, a velocity field $\vec{v}$ and embedded in a magnetic field $\vec{B}$. We write the mass conservation equation and the ideal steady momentum conservation equation as
\begin{align}
	\vec{\nabla}\cdot \left( \rho \vec{v}\right) = 0 \,, \label{equid1} \\
	\rho \vec{v}\cdot \vec{\nabla}\vec{v} = 
	- \vec{\nabla} \left(p + \frac{B^2}{2\muz}\right) 
	+ \frac{1}{\muz}\vec{B}\cdot \vec{\nabla}\vec{B} \,, \label{equid2}
\end{align}
where in the latter we separated the contributions of the magnetic pressure and tension. 
Combining together the electron force balance, $\vec{E} = \vec{v}\,\wedge\vec{B}$, with the irrotational character of the static electric field, $\vec{\nabla}\wedge\vec{E}=0$, we get the equation
\begin{equation}
	\vec{\nabla}\wedge \left( \vec{v}\,\wedge \vec{B}\right) = 0 \,. \label{equid3}
\end{equation}
Finally, the temperature $T$ verifies the stationary equation
\begin{equation}
	\vec{v}\cdot \vec{\nabla}T 
	+ \frac{2}{3}T\vec{\nabla}\cdot \vec{v} = 0 \,. \label{equid4}
\end{equation}
This configurational set is sufficient to properly characterize a plasma equilibrium, which we will specialize to the axial symmetry. 
\subsection{Axisymmetric equilibrium}\label{sec1.1}
We now consider axial symmetry in cylindrical coordinates $\{ r,\phi ,z\}$, so that all the partial derivatives $\partial_{\phi}(...) = 0$ vanish. Under this hypothesis, Eq.(\ref{equid1}) has the form
\begin{equation}
	\frac{1}{r}\partial_r\left(\rho rv_r\right) 
	+ \partial_z\left(\rho v_z\right) = 0 \,, \label{equid5}
\end{equation}
which admits the solution 
\begin{equation}
	\rho\vec{v} = -\frac{1}{r}\partial_z\Theta \hat{e}_r 
	+ \rho\omega r\hat{e}_{\phi} 
	+ \frac{1}{r}\partial_r\Theta \hat{e}_z \,, \label{equid6}
\end{equation}
$\Theta$ and $\omega$ being generic functions, the latter denoting the angular velocity of the plasma. We now consider a magnetic field having the form
\begin{equation}
	\vec{B} = -\frac{1}{r}\partial_z\psi \,\hat{e}_r 
	+ rK(\psi ) \,\hat{e}_{\phi} 
	+ \frac{1}{r}\partial_r\psi \,\hat{e}_z \,, \label{equid7}
\end{equation}
where $\psi$ is the magnetic flux function (times $1/2\pi$) and $K$ denotes a generic function. Since, in axial symmetry, the steady toroidal component of the electric field $E_{\phi}$ must vanish, then according to Eqs.(\ref{equid6}) and (\ref{equid7}) we get
\begin{equation}
	E_{\phi} = - \frac{1}{\rho r^2}\left( \partial_z\Theta\partial_r\psi - \partial_r\Theta\partial_z\psi \right) = 0 \,, \label{equid8}
\end{equation}
which implies the relation $\Theta = \Theta (\psi)$. Hence, comparing Eqs.(\ref{equid6}) and (\ref{equid7}), we immediately see the validity of the following relation: 
\begin{equation}
	\rho \vec{v}_p\equiv \rho \left( v_r\hat{e}_r + v_z\hat{e}_z\right) = \frac{d\Theta}{d\psi}\vec{B}_p \,, \label{equid9}
\end{equation}
where $\vec{v}_p$ and $\vec{B}_p$ denote the poloidal components of the velocity and the magnetic field, respectively. We now adopt the simplifying assumption $\rho = \rho (\psi )$, which implies, according to Eq.(\ref{equid1}), the two basic relations
\begin{equation}
	\vec{v}_p\cdot \vec{\nabla}\rho = 0 \, ; \quad \vec{\nabla}\cdot \vec{v} = 0 \,. \label{equid10}
\end{equation}
It is immediate to recognize that we also have
\begin{equation}
	\rho \vec{v}_p\cdot \vec{\nabla}\vec{v}_p =
	\frac{1}{\rho}\left(\frac{d\Theta}{d\psi}\right)^2 \vec{B}_p\cdot \vec{\nabla}\vec{B}_p \,. \label{equid11}
\end{equation}
Thus, the poloidal magnetic tension can be cancelled in Eq.(\ref{equid2}), by imposing the condition
\begin{equation}
	\frac{d\Theta}{d\psi} = 
	\pm \sqrt{\frac{\rho (\psi )}{\muz}} \,. \label{equid12}
\end{equation}
This equation also allows to cancel from the steady momentum equation the nonlinear poloidal advection term. Eq.(\ref{equid4}) for the plasma temperature, taking (\ref{equid10}) into account, gives
\begin{equation}
	\vec{v}_p\cdot \vec{\nabla}T = 0 \, \Rightarrow T = T(\psi ) \,. \label{equid13}
\end{equation}
Analogously, the toroidal component of Eq.(\ref{equid3}) (the poloidal one is now an identity) yields
\begin{equation}
	\partial_r\omega \partial_z\psi - \partial _z\omega\partial_r\psi = 0 \, \Rightarrow \omega = \omega (\psi) \,, \label{equid14}
\end{equation}
which expresses the corotation theorem \cite{ferraro37}.

\subsection{Momentum conservation equation}\label{sec2}
Now we analyze the momentum conservation equation (\ref{equid2}), starting from its toroidal component:
\begin{equation}
	\partial_z\psi \left(\mp\frac{\omega}{r}\sqrt{\frac{\rho}{\muz}}+\frac{K}{\muz r}\right) =
	-\rho\omega^2 r+\frac{rK^2}{\muz} \,, \label{equid15}
\end{equation}
where we used Eq.(\ref{equid12}). Then we find the simple relation for $K(\psi)$:
\begin{equation}
	K(\psi ) = \muz \omega \frac{d\Theta}{d\psi} = \pm \omega \sqrt{\muz \rho} \,. \label{equid16}
\end{equation}
Finally, we study the poloidal component of Eq.(\ref{equid2}), that is
\begin{equation}
	-r\left(\rho\omega^2-\frac{2K^2}{\muz} \right) \hat{e}_r = 
	-\vec{\nabla}\left( p + \frac{B_p^2}{2\muz}\right) 
	- \frac{r^2}{2\muz}\frac{dK^2}{d\psi}\vec{\nabla}\psi \,. \label{equid17}
\end{equation}
which, after some algebra and making use of Eq.(\ref{equid16}), can be rewritten in the following convenient form:
\begin{equation}
	\vec{\nabla}\left( p + \frac{B_p^2}{2\muz} + \rho\frac{\omega^2r^2}{2} \right) = 0 \,. \label{equid18}
\end{equation}
This equation clearly implies the existence of a constant quantity equal to the expression in brackets, which we label $E$. Its value can be determined considering the magnetic axis, \emph{i.e.} the point $A(r_A,z_A)$ where the magnetic flux function takes its maximum value, and its derivatives vanish. Hence we write the expression for $E$, along with the final equation for the plasma equilibrium, as:
\begin{align}
	E \equiv p_A + \rho_A \frac{\omega_A^2r_A^2}{2}\,, \label{equid19} \\
	\left(\partial_r\psi\right)^2 +\left(\partial_z\psi\right)^2 = 
	2 \muz r^2 \left[ E - p(r,z) 
	- \rho \frac{\omega^2 r^2}{2} \right] \,, \label{equid20}
\end{align}
where we introduced the notation $f_A=f(r_A,z_A)$ for the quantities calculated along the magnetic axis. Eq.(\ref{equid20}) is the configurational equation which allows to determine the equilibrium features of the considered plasma in the presence of matter flows.

\section{Tokamak relevant solution}\label{sec3}
We now search for a solution of the equilibrium which can be relevant in the context of a tokamak device. As a first step, in order to work with dimensionless variables, let us introduce the following normalizations and definitions:
\begin{align}
	r_n = \frac{r}{r_A} \,,\,\, z_n = \frac{z}{r_A} \,,\,\, \omega_n = \frac{\omega}{\omega_A} \,,\,\, 
	\rho_n = \frac{\rho}{\rho_A} \,,\,\, B_n = \frac{B}{B_A} \,, \nonumber \\
	\psi_n = \frac{\psi}{B_A r_A^2} \,,\,\, 	p_n = 2\muz \frac{p}{B_A^2} \,,\,\, 
	\chi_A \equiv \frac{\muz \rho_A \omega_A^2 r_A^2}{B_A^2} \,. \nonumber 
\end{align}
Using these variables, and dropping all subscripts $n$ from the notation for simplicity, Eq.(\ref{equid20}) can be recast as:
\begin{equation}
	\left(\partial_r\psi\right)^2 +\left(\partial_z\psi\right)^2 = 
	r^2 \left[ p_A - p(r,z) 
	+ \chi_A (1 - \rho \omega^2 r^2 ) \right] \,, \label{equid21}
\end{equation}

The usual approach to solving the Grad--Shafranov equation analytically is to assign a specific functional form for the source terms on the right--hand side (here $p$, $\rho$ and $\omega$), and then to solve the differential equation for the unknown function $\psi$. This method is natural when the equilibrium equation is linear in the magnetic flux $\psi$, and the source terms are functions of $\psi$ itself, allowing for suitable linearizations of the equation. In the present analysis, neither of these conditions hold; therefore, we look for a more convenient procedure, appropriate to evaluate the influence of the velocity field on the pressure profile. If we choose the pressure $p$ as unknown function, we can rewrite Eq.(\ref{equid21}) as:
\begin{equation}
	p(r,z) = p_A + \chi_A (1 - \rho \omega^2 r^2 )
	- \frac{\left(\partial_r\psi\right)^2 +\left(\partial_z\psi\right)^2}{r^2} \,, \label{equid22}
\end{equation}
Now if the magnetic flux function is known, we can compute its derivatives and plug them in the right--hand side; then, after assuming a reasonable functional form for $\rho$ and $\omega$, the pressure is readily obtained and its morphology can be studied. One way to obtain the function $\psi$ is to solve an equivalent equilibrium problem in the absence of plasma motion. In particular, we consider the well--known Solov'ev configuration, governed by
\begin{equation}
\Delta^*\psi + S_1 r^2 + S_2 = 0 \,, \label{equid23}
\end{equation}
where $S_1$ and $S_2$ are some constants linked to the plasma pressure and toroidal magnetic field \cite{solovev}. Hence we can write $\psi = \psi_0 - S_1 r^4/8 - S_2 z^2/2$, \emph{i.e.} the sum of the homogeneous and a particular solution, where $\psi_0$ satisfies $\Delta^*\psi_0 = 0$. Let us consider the following homogeneous solution for the up--down symmetric case \cite{egse,dpmarxiv}:
\begin{equation}
\psi_0(r,z) = r \sum_{i=1}^N \left[c_i I_1(r k_i) + d_i K_1(r k_i) \right] \cos(k_i z)\,,
\end{equation}
where $I_1$, $K_1$ denote the modified Bessel functions of order 1. $\{k_i\}$ is a set of $N$ arbitrary wavenumbers, which can roughly be taken of the order $\pi/\kappa$, $\kappa$ being the elongation of the plasma configuration. Then, the constants $c_i$, $d_i$ are determined by assigning a sufficient number of boundary points along the plasma magnetic separatrix, where $\psi=\psi_B$ (some constant value), and solving the resulting set of algebraic equations (this step is performed through a dedicated Mathematica code).

Once the function $\psi$ is known, we need to fix the dependence of the plasma density and rotation frequency. At this point, a consideration is important. According to Eq.(\ref{equid16}), the toroidal magnetic field $B_{\phi} = rK(\psi)$ is proportional to $\sqrt{\rho} \omega$. In tokamaks, the plasma equilibrium solution within the separatrix is necessarily linked to the vacuum solution of the surrounding region (the scrape--off layer) by matching boundary conditions. In the vacuum region, the magnetic field is essentially equal to the one generated by the external machine coils, thus it is generally different from zero. We conclude that in the present scenario $\rho$ and $\omega$ cannot vanish on the plasma magnetic separatrix, since this would imply a discontinuous geometry of the magnetic field lines. In other words, our solution is associated with density and rotation profiles which are non vanishing along the magnetic separatrix: the details related to such a discontinuous behaviour are not investigated further here, although they can be taken as a starting point to study H-mode plasma regimes, which are usually characterized by profiles with pedestals in proximity of the separatrix.

\begin{figure}[ht]
\centering
\begin{minipage}[b]{0.48\linewidth}
\includegraphics[width=0.91\linewidth]{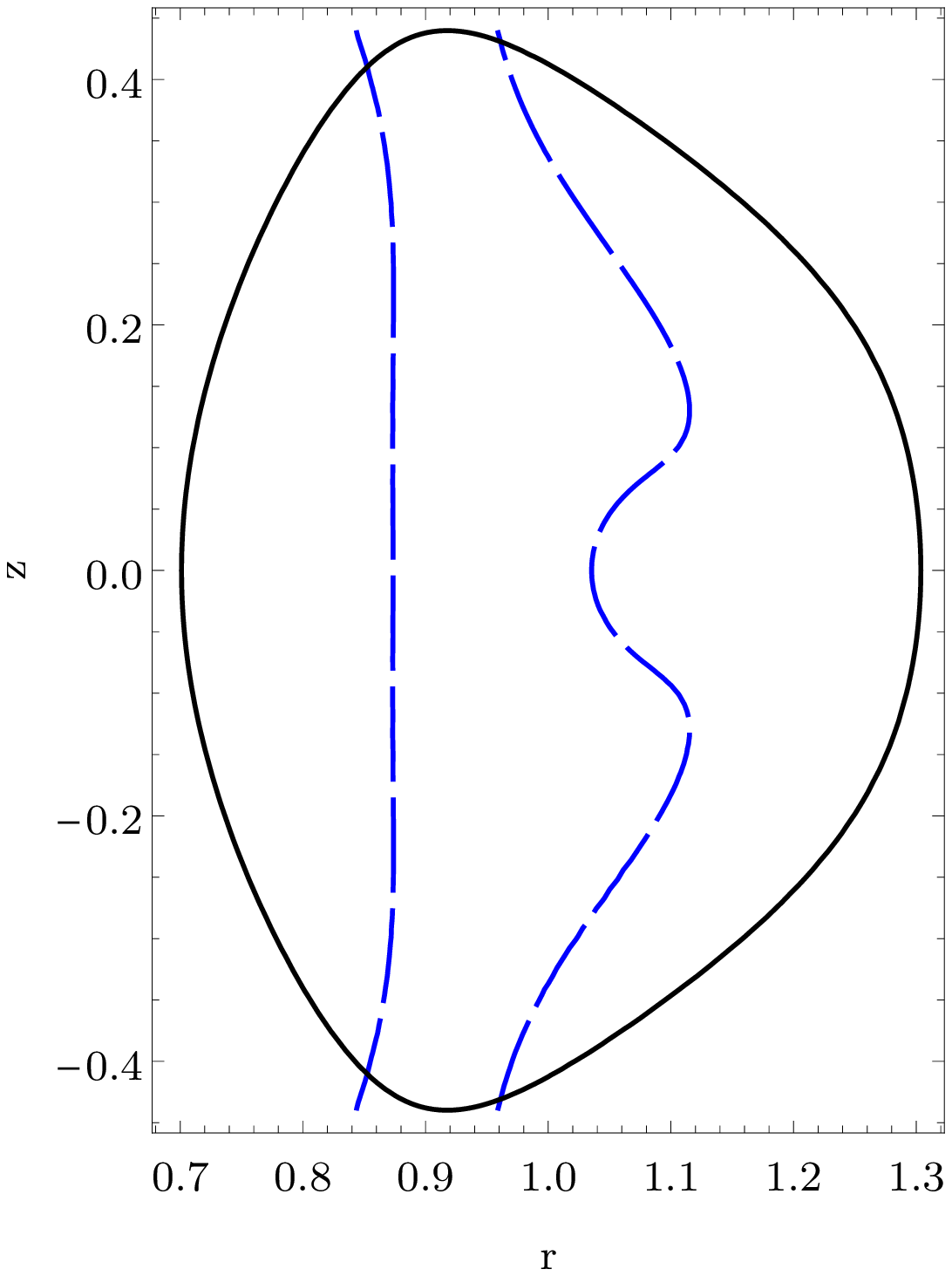}\vspace{-1pt}
\end{minipage}
\quad
\begin{minipage}[b]{0.45\linewidth}
\includegraphics[width=0.9\linewidth]{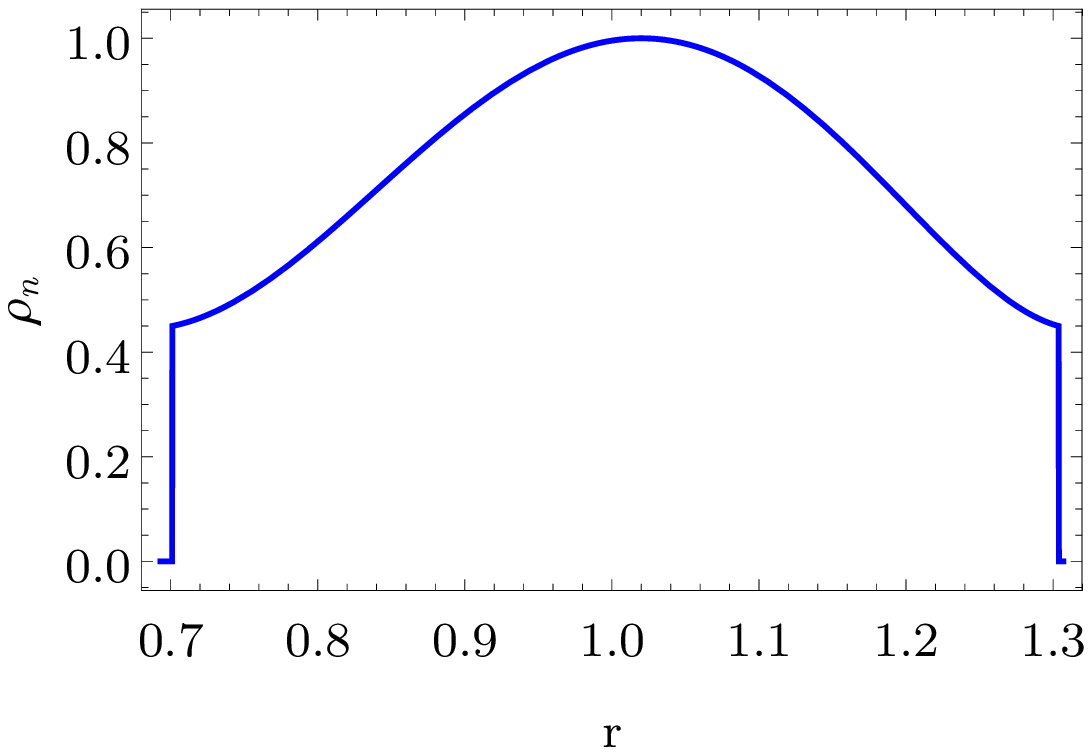}\vspace{5pt}\\
\includegraphics[width=0.9\linewidth]{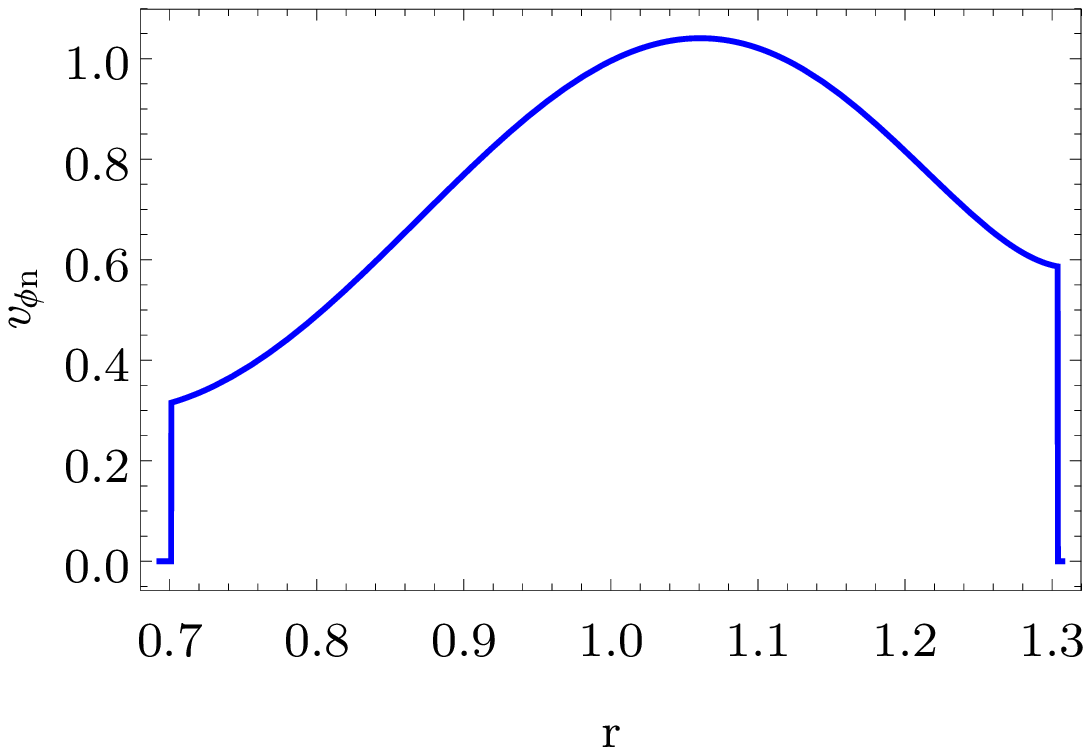}
\end{minipage}
\caption{(Left) Plasma magnetic separatrix (solid black) and isobar of null pressure (dashed blue) of a generic limiter plasma scenario with triangularity $\delta=0.35$. (Right) Profiles of normalized plasma density $\rho_n$ (top) and rotation velocity $r_n \omega_n$ (bottom) along the plasma equator.}
\label{fig1}
\end{figure}

\begin{figure}[ht]
\centering
\begin{minipage}[b]{0.48\linewidth}
\includegraphics[width=0.91\linewidth]{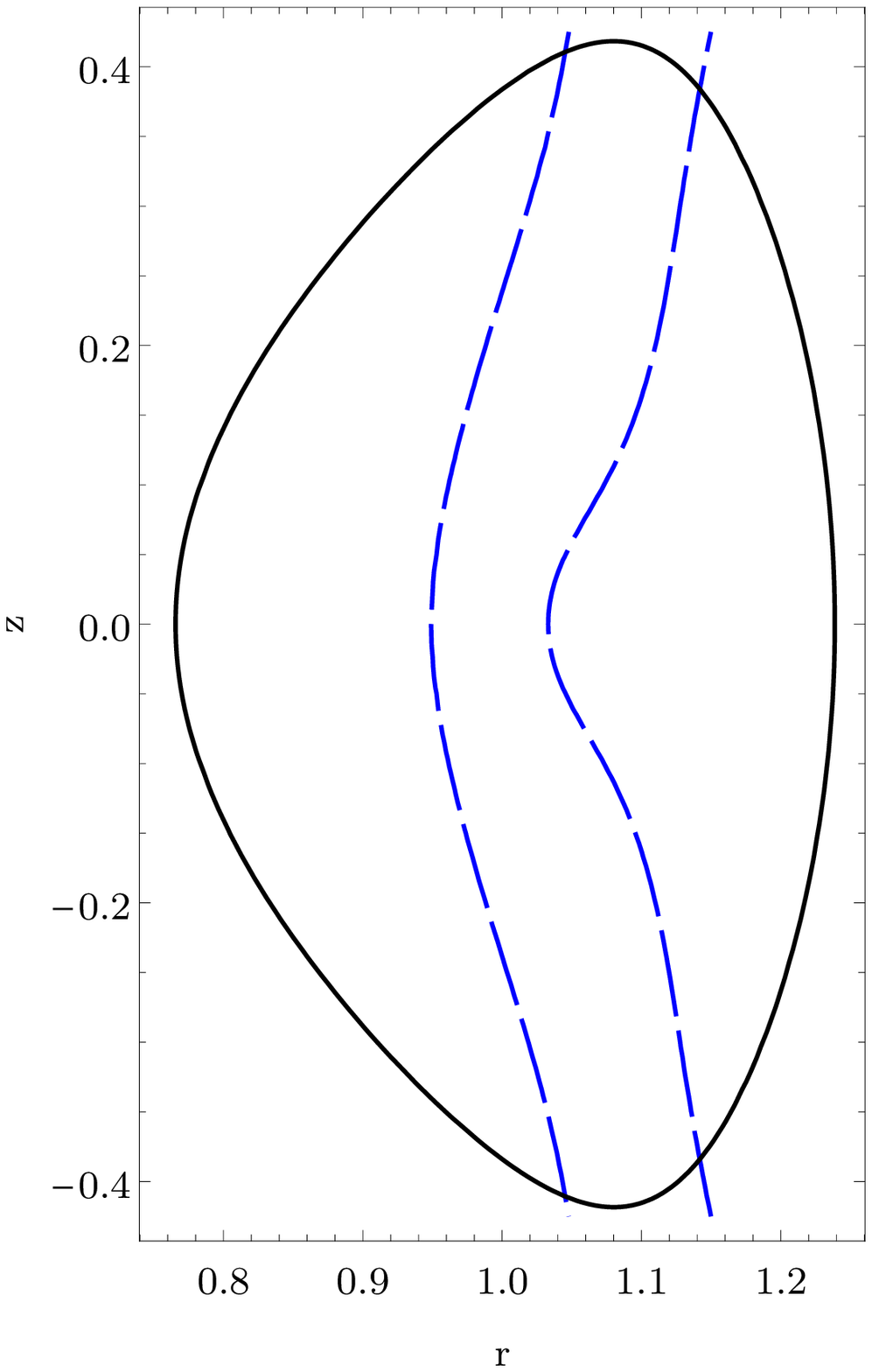}\vspace{-1pt}
\end{minipage}
\quad
\begin{minipage}[b]{0.45\linewidth}
\includegraphics[width=0.9\linewidth]{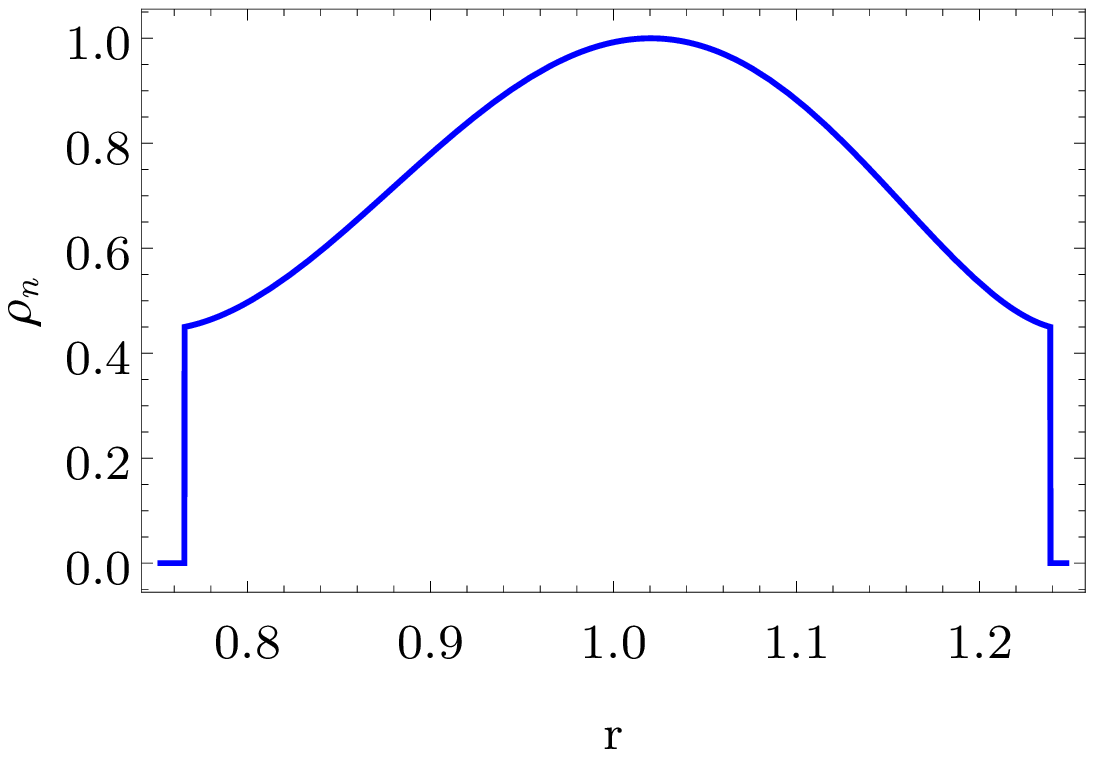}\vspace{5pt}\\
\includegraphics[width=0.9\linewidth]{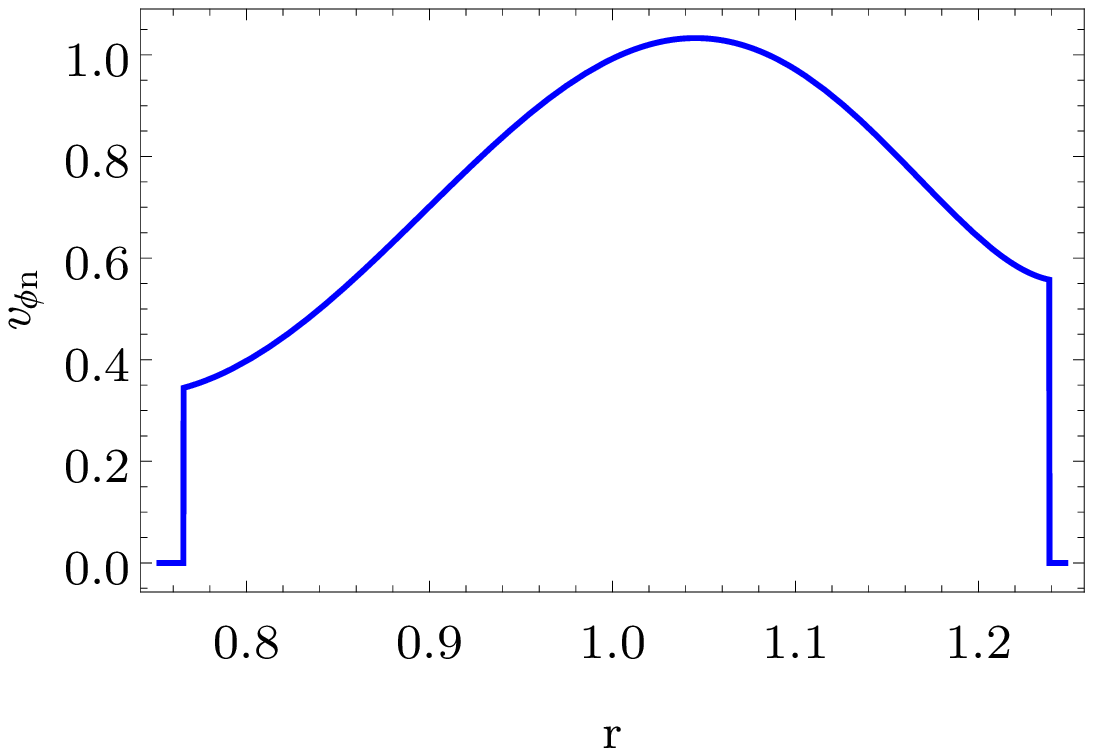}
\end{minipage}
\caption{(Left) Plasma magnetic separatrix (solid black) and isobar of null pressure (dashed blue) of a generic limiter plasma scenario with triangularity $\delta=-0.35$. (Right) Profiles of normalized plasma density $\rho_n$ (top) and rotation velocity $r_n \omega_n$ (bottom) along the plasma equator.}
\label{fig2}
\end{figure}

Let us consider the following normalized expressions:
\begin{equation}
\rho(\psi) = \frac{\rho_0 + \psi^2}{\rho_0 + 1} \,,\,\, \omega(\psi) = \frac{\omega_0 + \psi^2}{\omega_0 + 1} \,, \label{equid24}
\end{equation}
where the constants $\rho_0$ and $\omega_0$ can be assigned to control the pedestal height. In order to introduce the main features of our solution, let us show two illustrative cases, where we reproduce two generic plasma configurations in the absence of X-points (limited plasma), characterized by positive and negative triangularity, respectively (Figs.\ref{fig1} and \ref{fig2}). We emphasize the expected behaviour of the separation between surfaces of constant $\psi$ and $p$, due to plasma motion.

\begin{figure}[ht]
\centering
\begin{minipage}[b]{0.48\linewidth}
\includegraphics[width=0.91\linewidth]{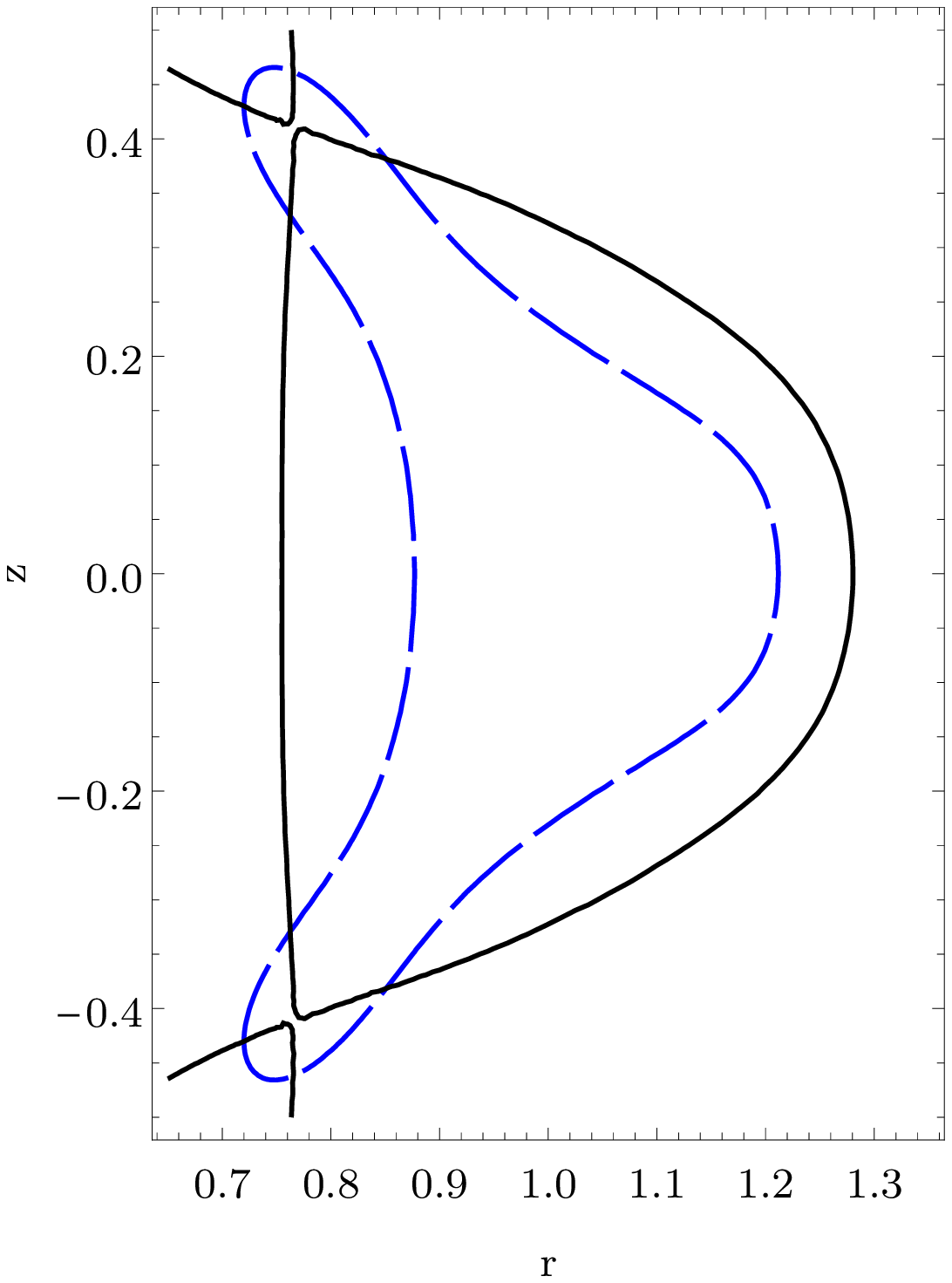}\vspace{-1pt}
\end{minipage}
\quad
\begin{minipage}[b]{0.45\linewidth}
\includegraphics[width=0.9\linewidth]{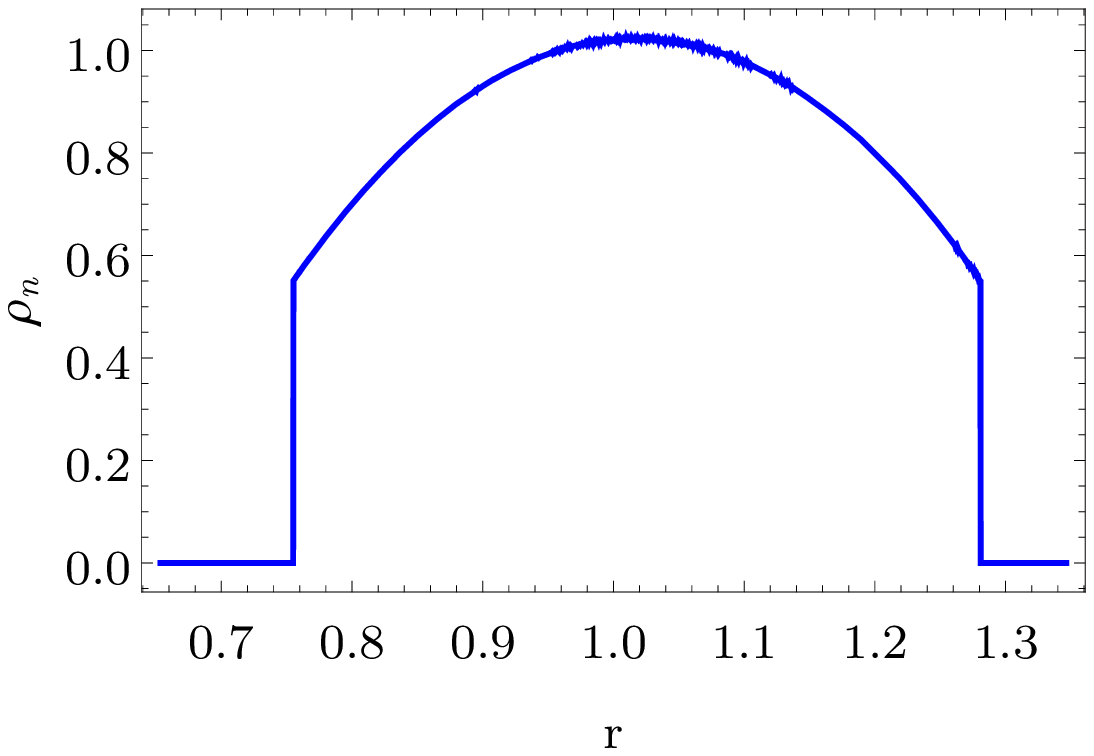}\vspace{5pt}\\
\includegraphics[width=0.9\linewidth]{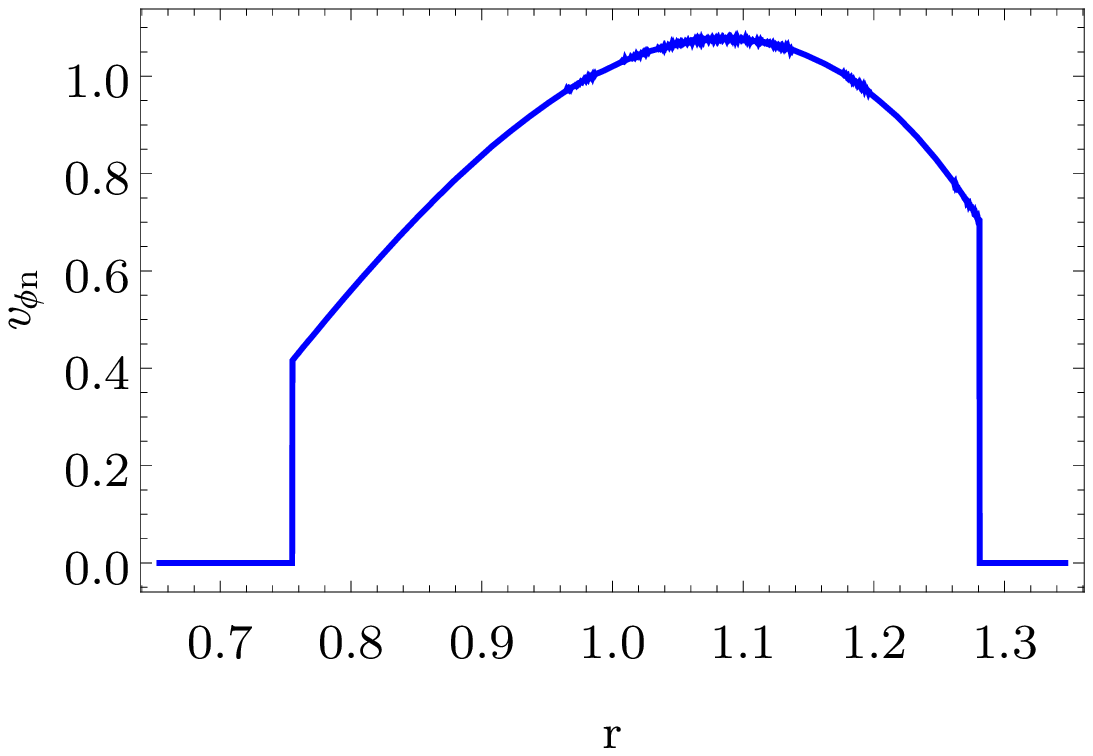}
\end{minipage}
\caption{(Left) Plasma magnetic separatrix (solid black) and isobar of null pressure (dashed blue) of a TCV-like equilibrium in the presence of two (symmetric) X-points. (Right) Profiles of normalized plasma density $\rho_n$ (top) and rotation velocity $r_n \omega_n$ (bottom) along the plasma equator.}
\label{fig3}
\end{figure}

The situation becomes interesting when we study a plasma scenario compatible with the TCV machine regime of operation \cite{tcvlike}. In particular, we assign a double--null magnetic geometry with high positive triangularity, with its X-points close to the machine first wall (Fig.\ref{fig3}).
Here the most striking feature is the behaviour of the curve of null pressure, which, departing from the magnetic separatrix, encircles the X-points, defining three distinct plasma confinement regions:
\begin{enumerate}
\item The region close to the plasma magnetic axis where $\psi > \psi_B$, $p > 0$ is qualitatively unaffected with respect to the static case, and corresponds to the usual plasma confinement condition.
\item The region inside the magnetic separatrix close the inner and outer plasma edges, where $\psi > \psi_B$ but outside of the positive pressure region, seemingly determining a deterioration of the plasma confinement due to the absence of a pressure gradient in the forces balance.
\item The region enclosed in the $p>0$ surface but outside of the magnetic separatrix, and especially its intersection with the ``private region'' of the plasma, that is defined by the divertor surface and the magnetic legs outside of the main plasma region.
\end{enumerate}
Concerning this latter case, our solution shows how, in this particular regime of toroidal and poloidal plasma motion, there is a confining effect due to the plasma pressure lines which is acting in the private region. Due to this enhanced confinement, the macroscopic effect would be that of relieving the particle and power exhaust of the plasma towards the divertor, with possible useful applications in upcoming devices where the management of the power output will be a main concern.

\section{Concluding remarks}
We constructed a new type of plasma equilibrium in axial symmetry, based on the assumption that the poloidal velocity and magnetic flux lines coincide along the configuration. Furthermore, we considered the plasma mass density only dependent on the magnetic flux function, basically assuming an incompressible plasma.

In this scenario, we arrived to a generalization of the Grad--Shafranov equation \cite{landau,grad,shafranov}, with the peculiar feature that it turns to be nonlinear in the gradient of the magnetic flux function. This can be explained by the fact that we were able to simultaneously cancel the nonlinear poloidal velocity term and the magnetic tension contribution. Thus, the resulting equilibrium equation contains the magnetic pressure only, from which arises the nonlinearity already mentioned. 

We then applied the obtained axisymmetric configuration to tokamak relevant situation and, as expected, we saw that the presence of velocity fields induces a deviation of the isobar surfaces (in particular that one at zero pressure) from the
morphology of the closed magnetic surfaces (in particular the separatrix). This fact is not surprising, but it was a relevant achievement to consider the implications of such a discrepancy between the separatrix and the zero pressure surface with respect to the presence of X-points in the magnetic profile.

In fact, analyzing configurations close to the operational conditions of the TCV device, we could observe a very interesting phenomenon, consisting in the inclusion of the X-points of a double--null scenario inside the surface of vanishing pressure. More specifically, such a surface envelopes these two symmetric (up and down) points, shielding the wall of the machine from power transfers to a degree which should be determined according to further studies. The private region is consequently divided into two parts, the one close to the divertor retains the usual morphology of the scrape--off layer, while the portion closer to the X-point acquires a pressure barrier since inside the surface $p=0$ the thermodynamical pressure supports the plasma confinement.

This result can impact the operation of a tokamak with diverted plasma, especially medium sized ones due to the presence of higher velocity fields, since the possibility to reach such configurations with "protected X-points'' could significantly reduce the heat and particle fluxes towards the divertor, favouring a better control of the power exhaust during the discharges.


\end{document}